\documentstyle[11pt,epsf,graphicx]{article}
\def\r{\mbox{\boldmath $r$}}

\def\p{\mbox{\boldmath $p$}}
\def\q{\mbox{\boldmath $q$}}
\begin{document}

\begin{center}
{\Large\bf{Nucleon-Nucleon Correlations in Electromagnetically Induced Knockout 
Reactions}}
\end{center}
\vskip1.0truecm
\centerline{\large{C. Giusti\footnote[1]{E-mail: giusti@pv.infn.it, 
phone: +39 0382507454, fax: +39 0382526938.}}}
\vskip1.0truecm
\centerline{\small{Dipartimento di Fisica Nucleare e Teorica, 
Universit\`a degli Studi di Pavia, and}}

\centerline{\small{Istituto Nazionale di Fisica Nucleare, Sezione di
Pavia, Pavia, Italy}}

\begin{abstract}
The attempts to investigate correlations in electromagnetically induced 
one- and two-nucleon knockout are reviewed. The theoretical framework for cross
section calculations is outlined and some results are presented for the 
exclusive $^{16}$O($e,e'p$)$^{15}$N and $^{16}$O($e,e'pp$)$^{14}$C  
reactions. For the ($e,e'p$) reaction attention is focussed on extracting the
spectroscopic factors. For the ($e,e'pp$) reaction the possibility of obtaining 
direct and clear information on short-range correlations is discussed.  
\end {abstract}
\vspace{1cm}

\section{Introduction}
Correlations in the nuclear wave function beyond the mean-field approximation
are very important to describe the basic properties of  nuclear structure. The
mean-field (MF) or Hartree-Fock (HF) approximation gives a good description of
the bulk properties of nuclei if phenomenological nucleon-nucleon ($NN$) forces
are used, but employing realistic $NN$ interactions it fails badly. The
failure is a consequence of the strong and repulsive short-range components of a
realistic $NN$ interaction. Thus, a careful evaluation of the short-range
correlations (SRC), produced by the short-range components of the $NN$
interaction, is needed to describe the nuclear properties in terms of a 
realistic $NN$ interaction and provide profound insight into the structure of 
the hadronic interaction in a nucleus. The same is true for the tensor 
correlations (TC), which are induced by the strong tensor components of the 
$NN$ interaction. Moreover, it is necessary to consider also those processes 
beyond the MF approximation falling under the generic name of long-range 
correlations (LRC), which are related to the coupling between the 
single-particle (s.p.) dynamics and the collective excitation modes of the 
nucleus and which mainly represent the interaction of nucleons at the nuclear 
surface. These processes can be very important in finite nuclear systems. A 
consistent evaluation of these different types of correlations is needed.

Various theoretical methods have been developed to account for such correlation
effects (see, e.g., \cite{Ant88,Ant93,MP}). Besides developing these methods, it 
has always been a challenge for nuclear physics to envisage experimental 
situations and observables, particularly sensitive to correlations, whose 
determination can give clear evidence for correlations, in particular for SRC. 
The hope is that the comparison between the predictions of different models and 
data can give detailed information on correlations and can allow one to 
distinguish the different models of the $NN$ interaction at short distance. 

Electromagnetically induced one- and two-nucleon knockout reactions are 
powerful tools for this study (see, e.g., \cite{Oxford}). In one-nucleon knockout 
the conditions of individual nucleons in the nuclear medium can be explored. 
For an exclusive reaction, the coincidence cross section contains the one-hole 
spectral density function, i.e. 
\begin{equation}
S(\p_1,\p^{\prime}_1;E_{\mathrm m})=\langle \Psi_{\mathrm{i}} |
a^+_{\p^{\prime}_1} \delta(E_{\mathrm m} -H)a_{\p_1}| \Psi_{\mathrm{i}} 
\rangle,
\label{eq:HSF}
\end{equation}
which in its diagonal form ($\p_1=\p^{\prime}_1$) gives the joint probability
of removing from the target a nucleon, with momentum $\p_1$, leaving the
residual nucleus in a state with energy $E_{\mathrm m}$ with respect to the
target ground state. In an inclusive reaction, integrating the spectral density 
over the whole energy spectrum produces the one-body density matrix (OBDM) 
$\rho(\p_1,\p^{\prime}_1)$, that in its diagonal form gives the nucleon 
momentum distribution. 

Likewise two-nucleon knockout represents the preferential tool for exploring the
conditions of pairs of nucleons in the nuclear medium. For an exclusive 
reaction, the coincidence cross section contains the two-hole spectral density 
function, i.e.
\begin{equation}
S(\p_1,\p_2,\p^{\prime}_1,\p^{\prime}_2;E_{\mathrm m})=
\langle \Psi_{\mathrm{i}} |a^+_{\p^{\prime}_2} a^+_{\p^{\prime}_1}
\delta(E_{\mathrm m} -H)a_{\p_1} a_{\p_2}| \Psi_{\mathrm{i}} \rangle,
\label{eq:H2SF}
\end{equation}
which in its diagonal form ($\p_1=\p^{\prime}_1$ and $\p_2=\p^{\prime}_2$) gives 
the joint probability of removing from the target two nucleons, with momenta 
$\p_1$ and $\p_2$, leaving the residual nucleus in a state with energy 
$E_{\mathrm m}$ with respect to the target ground state. In an inclusive
reaction, integrating the spectral density over the whole energy spectrum 
produces the two-body density matrix 
$\rho(\p_1,\p_2,\p^{\prime}_1,\p^{\prime}_2)$, that in its diagonal form
and in the coordinate representation gives the pair correlation function, i.e. 
the conditional probability density of finding in the target a particle at 
$\r_2$ if another one is known to be at $\r_1$.

The cross sections of one- and two-nucleon knockout reactions contain 
information on $NN$ correlations in the spectral functions and in the 
density matrices. In order to extract this information, along with the 
experimental work, a reliable model for cross section calculations 
is needed, able to keep reasonably under control the reaction mechanism and all 
the theoretical ingredients contributing to the cross section. 

The quasifree ($e,e'p$) reaction has extensively been used to investigate the 
s.p. properties of nuclei and to point out the validity and the limits of the 
independent-particle shell model (IPSM) \cite{Oxford}. The fact that a pure MF 
picture is unable to give a precise quantitative description of ($e,e'p$) data 
is due to correlations and the discrepancy can give a measurement of 
correlation effects. More direct information on correlations, and in particular 
on SRC, is available from two-nucleon knockout reactions \cite{Oxford}. First 
recent experiments have given clear evidence for SRC in the 
$^{16}$O($e,e'pp$)$^{14}$C reaction \cite{Ronald}. 

In this contribution the attemps to investigate $NN$ correlations in exclusive 
knockout reactions are reviewed. The theoretical framework for cross section 
calculations is outlined and some results are presented and discussed for 
($e,e'p$) in Sec. 2 and ($e,e'pp$) in Sec. 3. 

\section{One-nucleon knockout}

\subsection{The Plane-Wave Impulse Approximation}

In the one-photon exchange approximation, where the incident electron exchanges 
a photon of momentum $\q$ and energy $\omega$ with the target, the most general 
form of the coincidence ($e,e'N$) cross section  involves the contraction 
between a lepton tensor $L_{\mu\nu}$ and a hadron tensor $W^{\mu\nu}$. The 
lepton tensor is produced by the matrix elements of the electron current and, 
neglecting the effect of the nuclear Coulomb field on electrons, contains only 
electron kinematics. The components of the hadron tensor are given by bilinear
combinations of the Fourier transforms of the transition matrix elements of the 
nuclear current operator between initial and final nuclear states, i.e.
\begin{equation}
J^{\mu}({\mbox{\boldmath $q$}}) = \int \langle\Psi_{\mathrm{f}} 
|\hat{J}^{\mu}({\mbox{\boldmath $r$}})|\Psi_{\mathrm{i}}\rangle
{\mathrm{e}}^{\,{\mathrm{i}}{\footnotesize {\mbox{\boldmath $q$}}}\cdot 
{\footnotesize {\mbox{\boldmath $r$}}}}
{\mathrm d}{\mbox{\boldmath $r$}}.     \label{eq:jm}
\end{equation} 

In the plane-wave impulse approximation (PWIA), i.e. neglecting final-state
interactions (FSI) of the ejected particle, the ($e,e'p$) cross section is
factorized as a product of a kinematical factor, the (off-shell) electron-proton
cross section and, for an exclusive reaction, the one-hole diagonal spectral 
density function 
\begin{equation}
S(\p_{\mathrm m},E_{\mathrm m})= \sum_{\alpha} S_\alpha(E_{\mathrm m})
|\phi_{\alpha}(\p_{\mathrm m})|^2,
\label{eq:pwia}
\end{equation}
where the missing momentum $\p_{\mathrm m}$ is the recoil momentum of the 
residual nucleus. At each value of $E_{\mathrm m}$, the momentum dependence of 
the spectral function is given by the momentum distribution of the quasi-hole 
states $\alpha$ produced in the target nucleus at that energy and described by 
the normalized overlap functions (OF) $\phi_\alpha$ between the target ground 
state and the states of the residual nucleus. The (normalization) spectroscopic 
factor (s.f.)  $S_\alpha$ gives the probability that the quasi-hole state
$\alpha$ is a pure hole-state in the target. In an IPSM $\phi_\alpha$ are the 
s.p. states of the model and $S_\alpha =1(0)$ for occupied (empty) states. In 
reality, the strength of a quasi-hole state is
fragmented over a set of s.p. states, and $0 \leq S_\alpha < 1$. The
fragmentation of the strength is due to correlations and the s.f. can thus give 
a measurement of correlation effects. 

The PWIA is a simple and clear picture that is able to describe the main
qualitative features of ($e,e'p$) cross sections, but is unable to give a 
precise quantitative description of data. For the analysis of data a more 
refined theoretical treatment is needed. The calculations for this analysis 
were carried out with the program DWEEPY \cite{DWEEPY}, within the theoretical 
framework of a nonrelativistic distorted-wave impulse approximation (DWIA), 
where FSI and Coulomb distortion of the electron wave functions are 
taken into account.

\subsection{The Distorted-Wave Impulse Approximation}

The DWIA treatment of the matrix elements in Eq.~(\ref{eq:jm}) is based on the
following assumptions :
\newline
i) An exclusive process is considered, where the residual nucleus is left in a
discrete eigenstate $|\Psi^{\mathrm B}_\alpha(E) \rangle$ of its Hamiltonian,
with energy $E$ and quantum numbers $\alpha$.
\newline
ii) The final nuclear state is projected onto the channel subspace spanned by 
the vectors corresponding to a nucleon, at $\r_1$, and the residual nucleus in 
the state $|\Psi^{\mathrm B}_\alpha(E) \rangle$. This assumption neglects 
effects of coupled channels and is justified by the considered asymptotic 
configuration of the final state.
\newline
iii) The (one-body) nuclear-current operator does not connect different channel 
subspaces. Thus, also the initial state is projected onto the selected channel 
subspace. This assumption is the basis of the direct knockout (DKO) mechanism 
and is related to the IA.

The transition matrix elements in Eq.~(\ref{eq:jm}) can thus be written in
a one-body representation as
\begin{equation}
J^{\mu}(\q) = \int \chi^{(-)*}_{E\alpha}(\r_1)\hat{J}^{\mu}(\r,\r_1) 
\phi_{E\alpha}(\r_1)\left[S_\alpha(E)\right]^{1/2}
{\mathrm{e}}^{\,{\mathrm{i}} {\footnotesize {\q}}\cdot
{\footnotesize {\r}}} {\mathrm d}\r {\mathrm d} \r_1 ,
\label{eq:dwia}
\end{equation}
where
\begin{equation}
\chi^{(-)}_{E\alpha}(\r_1) =
\langle\Psi^{\mathrm B}_\alpha(E)|a_{\r_1}| \Psi_{\mathrm f}\rangle
\label{eq:dw}
\end{equation}
is the s.p. distorted wave function of the ejectile and the overlap function
\begin{equation}
\left[S_\alpha(E)\right]^{1/2}\phi_{E\alpha}(\r_1) =
\langle\Psi^{\mathrm B}_\alpha(E)|a_{\r_1} |\Psi_{\mathrm i}\rangle
\label{eq:ovf}
\end{equation}
describes the residual nucleus as a hole state in the target. The
spectroscopic strength $S_\alpha(E)$ is the norm
of the overlap integral in the right-hand side of (\ref{eq:ovf}) and gives the
probability of removing from the target a nucleon at $\r_1$ leaving the residual 
nucleus in the state $|\Psi^{\mathrm B}_\alpha(E) \rangle$.

The scattering state in Eq.~(\ref{eq:dw}) and the normalized bound state 
$\phi_{E\alpha}(\r_1)$ in Eq.~(\ref{eq:ovf}) are consistently derived in this 
model from an energy-dependent non-Hermitean optical model Feshbach Hamiltonian.
In standard DWIA calculations, however, phenomenological ingredients are 
employed. The nucleon scattering state is eigenfunction of a phenomenological 
optical potential, determined through a fit to elastic nucleon-nucleus 
scattering data including cross sections and polarizations. Phenomenological 
bound-state wave functions are usually adopted for the OF, which thus do not 
include correlations. In the analysis of data these functions were calculated 
in a Woods-Saxon well, where the radius was determined to fit the experimental 
momentum distributions and the depth was adjusted to reproduce the 
experimentally observed separation energy of the bound final state. The 
normalization of the wave function was fitted to the data. In order to 
reproduce the magnitude of the experimental cross sections, a reduction factor 
was applied to the calculated results. This factor was then identified with the 
s.f.
 
The ``experimental'' s.f. extracted in these DWIA analyses indicate that the 
removal of the s.p. strength for quasi-hole states near the Fermi energy is 
about 60-70\%. The s.f. gives a measurement of correlation effects, but, since 
in the ($e,e'p$) analyses it is obtained through a fit to the data, in practice 
it can include besides correlations also the effect of other contributions 
which are neglected or not adequately described in the model. It 
can be identified with the s.f. only if all the theoretical ingredients 
contributing to the cross section are reasonably under control. On the other 
hand, the fact that this model, with phenomenological ingredients, was able to 
give an excellent description of ($e,e'p$) data, in a wide range of nuclei and 
in different kinematics (see, e.g., \cite{Oxford,Lap93}), gives support and 
consistency to this whole picture and to the interpretation of the s.f. 
extracted in comparison with data.   

\subsection{Overlap Functions and Correlations}

Explicit calculations of the hole spectral function and the associated fully 
correlated OF for complex nuclei are very difficult. Only recently the first 
successful parameter-free comparison of experiment and theory including the 
absolute normalization in $p$-shell nuclei has been performed for the 
$^7$Li($e,e'p$) reaction \cite{Li7}. For heavier nuclei a calculation able to 
account for the effects due to all types of correlations appears extremely 
difficult, since it requires excessively large model space. The effects of a 
spectral function containing only SRC and TC \cite{PR} and only LRC 
\cite{Amir} have been investigated in the $^{16}$O($e,e'p$) reaction. A method 
to deal with SRC and LRC consistently has been proposed and applied in 
ref. \cite{Geurts} to $^{16}$O. In this application, however, only the s.f. and 
not the OF have been calculated.

Recently, a general procedure has been adopted \cite{Vn93} to extract the 
OF and the associated s.f. on the base of the OBDM. The advantage of this 
procedure is that it avoids the complicated task of calculating the nuclear 
spectral function, but its success depends on the availability of realistic 
calculations of the OBDM. 

This procedure has been applied \cite{Sto96,Di97,Gai99,Gai2000,Iva2001} to OBDM
of $^{16}$O and $^{40}$Ca constructed within different correlation methods, 
such as the Jastrow Correlation Method (JCM) \cite{Sto96}, the Correlated Basis 
Function (CBF) theory \cite{Vn97,Sa96}, the Green's Function Method 
(GFM) \cite{Po96}, and the  Generator Coordinate Method (GCM) 
\cite{Chr86,Iva2000}. The OF and the s.f. have then been used to calculate the 
cross section of one-nucleon removal reactions \cite{Gai99,Gai2000,Iva2001}. 

\begin{figure}[ht]
\begin{center}
\includegraphics[height=100mm,width=100mm]{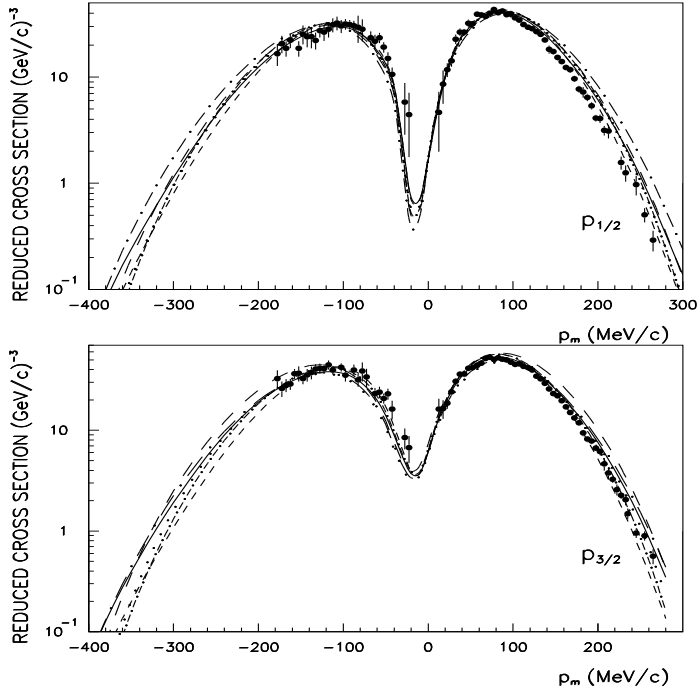}
\vspace{2mm}\caption{Reduced cross sections of  the
$^{16}$O($e,e'p$) reaction as a function of the missing momentum 
$p_{\mathrm m}$ for the transitions to the $1/2^{-}$ ground state and to the 
first $3/2^{-}$ excited state of $^{15}$N in parallel kinematics, with an 
incident electron energy $E_0=520.6$ MeV and an outgoing proton energy 
$T'_1=90$ MeV. 
The optical potential is from ref. [24]. The OF are derived from 
the OBDM of GFM [20] (solid line), CBF [18] (long-dashed line), 
CBF [19] (long-dot-dashed line), JCM [13] (short-dot-dashed 
line), and GCM [21] (short-dashed line). The dotted line is calculated 
with the HF wave function. The positive (negative) values of $p_{\mathrm m}$ 
refer to situations where $|\q|<|\p'|$ ($|\q|>|\p'|$). The experimental data 
are taken from ref. [23]. The theoretical results
have been multiplied by the reduction factor given in column II of Table I
(from ref. [16]).
\label{fig:fig1}
}
\end{center}
\end{figure}

An example is displayed in Fig.~\ref{fig:fig1} for the $^{16}$O($e,e'p$) 
reaction \cite{Gai2000}. Calculations have been done with same code 
DWEEPY \cite{DWEEPY} used 
in the original analysis of the NIKHEF data \cite{Leuschner}, in the same 
conditions and with the same optical potential \cite{Schwandt}, but the 
phenomenological s.p. bound state wave functions have been replaced by the 
theoretically calculated OF. The results are presented in terms of the reduced 
cross section \cite{Oxford}, defined as the cross section divided by a 
kinematical factor and the elementary off-shell electron-proton scattering cross 
section, which is the quantity that in PWIA gives the momentum distribution of 
the quasi-hole state. The experimental data were taken in the so-called parallel 
kinematics, where the momentum of the outgoing nucleon is fixed and is taken 
parallel or antiparallel to the momentum transfer. Different values of 
$\p_{\mathrm m}$ are obtained by varying the electron scattering angle and 
therefore the magnitude of the momentum transfer. 
In Fig.~\ref{fig:fig1} the calculated reduced cross sections are able to 
reproduce with a fair agreement the experimental distributions. They are anyhow 
sensitive to the shape of the various functions. The differences are larger at 
large values of $p_{\mathrm m}$, where correlation effects are more sizable. 
The best agreement with data, for both transitions, is obtained with the OF 
emerging from the OBDM calculated within the GFM \cite{Po96} and corresponding 
to the most refined calculation of the OBDM.

The results obtained with the different OF are compared in the figure with 
those given by the HF wave function, which is calculated in a self-consistent 
way using the Skyrme-III interaction. Besides the HF wave function, whose norm 
is equal to one, all the  OF contain a s.f. These factors are listed in 
Table I (column I). They account for the contribution of correlations included 
in the OBDM, which cause a depletion of the quasi-hole states. Only SRC are 
included in the OBDM of refs. \cite{Sto96,Sa96,Chr86}, whereas also TC are 
taken into account in refs. \cite{Vn97,Po96}. Indeed the s.f. are lower for the 
functions including also TC. These OF, however, do not include LRC, which 
should produce further depletion of the quasi-hole states \cite{Amir,Geurts}. 
In order to reproduce the size of the experimental cross section, a reduction 
factor has been applied in Fig.~\ref{fig:fig1} to the calculated results. 
These factors are also listed in Table I (column II). They can be considered 
as additional s.f. reflecting the depletion of the quasi-hole state produced by 
the correlations not included in the OBDM, namely LRC for the OBDM of 
refs. \cite{Vn97,Po96}, LRC and TC for those of refs. \cite{Sto96,Sa96,Chr86}, 
and all the correlations for the HF wave function. The product of the two 
factors, in column III, can thus be considered as the total s.f. accounting 
for the combined effect of all the correlations. Indeed for $1p_{1/2}$ these 
factors are in reasonable agreement with the s.f. (0.77) obtained in the 
calculation of ref. \cite{Geurts}, where both SRC and LRC are consistently 
included. The fact that for $1p_{3/2}$ these factors are lower than the one 
found in \cite{Geurts} (0.76) is presumably due to the approximations used 
in that calculation, which is unable to reproduce the experimentally observed 
splitting of the $3/2^{-}$ state. Further work is currently in progress to 
account for the complexity of the low-energy structure of $^{16}$O \cite {BD}. 

\begin{figure}[ht]
\begin{center}
\includegraphics[height=60mm,width=120mm]{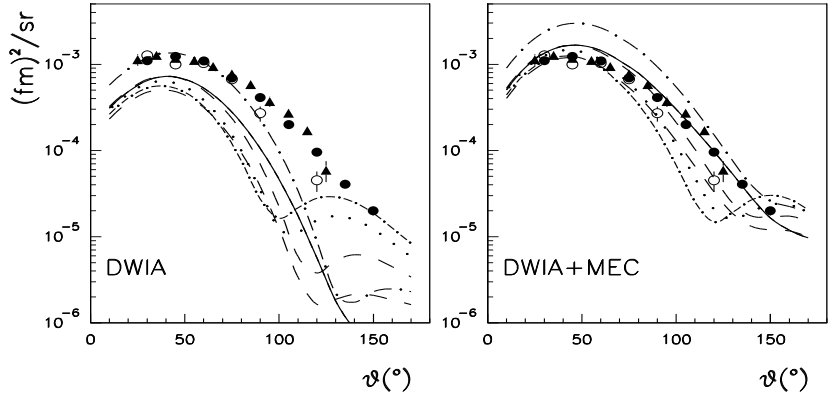}
\vspace{2mm}\caption{Angular distribution of the cross section of the
$^{16}$O($\gamma,p$) reaction for the transition to the $1/2^{-}$ ground 
state of $^{15}$N  at $E_\gamma = 60$ MeV. The separate contribution given by 
the one-body current (DWIA) and the final result given by the sum of the 
one-body and the two-body seagull current (DWIA+MEC) are shown. Line convention 
as in Fig.~1. The optical potential is from ref. [24]. 
The experimental data are taken from refs. [26] (black circles), 
[27] (open circles) and [28] (triangles). The theoretical 
results have been multiplied by the reduction factors listed in column II of 
Table I, consistently with the analysis of ($e,e'p$) data 
(from ref. [16]).
\label{fig:fig2}
}
\end{center}
\end{figure}

The DWIA calculations with the different OF for the 
$^{16}$O($\gamma,p$)$^{15}$N$_{\mathrm{g.s.}}$ reaction at $E_\gamma = 60$  MeV 
are displayed in  Fig.~\ref{fig:fig2} \cite{Gai2000}. The same theoretical 
ingredients, i.e. OF, s.f., and consistent optical potentials, have been 
adopted as in ($e,e'p$). Moreover, the reduction factor determined in 
comparison with the ($e,e'p$) data has been applied, in order to allow a 
consistent comparison of ($e,e'p$) and ($\gamma,p$) results. In photon-induced 
reactions a different kinematics is explored. In fact in this case the energy 
and momentum transfer cannot be independently varied. They are constrained by 
the condition $\omega=|\q|=E_\gamma$, and only the high-momentum components of 
the nuclear wave function are probed, higher values than in the usual kinematics 
of ($e,e'p$) experiments. Thus, it is not strange that the differences given by 
the different OF are so large, in particular at backward angles, which 
correspond to higher values of $p_{\mathrm m}$. The agreement of the DWIA 
calculations with data is poor. For the ($\gamma,p$) reaction the validity of 
the DKO mechanism related to the DWIA, which is clearly stated for ($e,e'p$), 
is much more questionable and large contributions are expected also by 
two-nucleon processes, such as those involving two-body meson-exchange currents 
(MEC). Indeed a better agreement with data is obtained in Fig.~\ref{fig:fig2}
when MEC are added. Here MEC have been added to the DKO mechanism within the 
theoretical framework of ref. \cite{Benenti}, in a microscopic and unfactorized 
calculation. In order to reduce the complexity of the calculation, however, only 
the contribution due to the seagull diagrams with one-pion exchange has been 
included. This is certainly an approximation, but the seagull current here 
considered should give the main contribution of the two-body current in the 
photon-energy range above the giant resonance and below the pion production 
threshold. MEC give an important contribution and bring the results closer to 
data. The differences for the various OF are however still large and for some 
functions the agreement with data is still poor. The best agreement and a fair 
description of data is given by the OF from GFM \cite{Po96}, which is able to 
give also the best description of ($e,e'p$) data and with the same s.f. This 
result, that has been confirmed also in different situations \cite{Gai2000}, is 
a strong indication in favour of a consistent description of the two reactions 
and gives further support to our results for ($e,e'p$), where the contribution 
to the depletion of the quasi-hole states produced by the different types of 
correlations has been established. A part of this contribution, however, is 
still obtained as a reduction factor in comparison with data. Thus, also in 
this analysis the s.f. can be affected by other effects not included or not 
adequately decribed in the theoretical model. 

\subsection{Spectroscopic Factors and Relativistic Effects}

Various contributions and their effect on the extracted s.f. have been studied 
in recent years. Only a small contribution is expected from two-body currents 
in ($e,e'p$) \cite{BR,VdS,ALC}. A proper treatment of the c.m. motion leads to 
an enhancement of the extracted s.f. by about 7\% \cite{CM}. A further 
enhancement is obtained in relativistic DWIA (RDWIA) analyses with relativistic 
optical potentials \cite{relat,darwin}.

Fully relativistic models based on the RDWIA have been developed by different
groups \cite{relat,Meucci}. In these approaches the bound 
nucleons are described by s.p. Dirac wave functions in the presence of scalar 
and vector potentials fitted to the ground-state properties of the nucleus, and 
the scattering wave function is solution of the Dirac equation with 
relativistic optical potentials obtained by fitting elastic proton-nucleus 
scattering data. Also RDWIA analyses are able to give a good description of
($e,e'p$) data. RDWIA calculations are necessary for the analyses of the new 
($e,e'p$) data from Jlab \cite{Gao} in kinematic conditions inaccessible in 
previous experiments, where the four-momentum transfer squared $Q^2$ was less 
than 0.4 (GeV/$c$)$^2$ and the outgoing proton energy generally around 100 MeV. 
It is anyhow important to check the relevance of relativistic effects also in the
kinematics at lower energies of the previous experiments, whose data were
analyzed with a nonrelativistic DWIA treatment. 

Relativistic effects as well as the differences between relativistic and
nonrelativistic calculations have been investigated in different papers where 
RDWIA treatments have been developed. The differences, however, are usually 
evaluated starting from the basis of a relativistic model where terms 
corresponding to relativistic effects are cancelled or where nonrelativistic 
approximations are included. Although very interesting, these investigations do 
not correspond to the result of a comparison between RDWIA and the DWIA 
calculations carried out with the program DWEEPY. In fact, DWEEPY is based on a 
nonrelativistic treatment where some relativistic corrections are introduced in 
the kinematics and in the nuclear current operator. On the other hand, only 
indirect comparison between relativistic and nonrelativistic calculations can 
be obtained from the available data analyses carried out with DWEEPY and in 
RDWIA. In fact, the two types of calculations make generally use of different 
optical potentials and bound state wave functions, and the difference due to 
the different theoretical ingredients cannot be attributed to relativity. 

In order to investigate the relevance of genuine relativistic effects through 
a direct comparison between RDWIA calculations and the results of DWEEPY, a 
fully relativistic RDWIA model for the ($e,e'p$) reaction has been developed 
and its numerical results have been compared with the corresponding results 
given by DWEEPY \cite{Meucci}. In order to make the comparison as consistent as 
possible, in the nonrelativistic calculations the bound state is the normalized 
upper component of the Dirac spinor and the scattering state is the solution of 
the same Schr\"odinger-equivalent optical potential of the relativistic 
calculation. This is not the best choice for DWEEPY, but the same theoretical 
ingredients are to be used for a clear comparison between the two approaches. 

\begin{figure}[ht]
\begin{center}
\includegraphics[height=100mm,width=100mm]{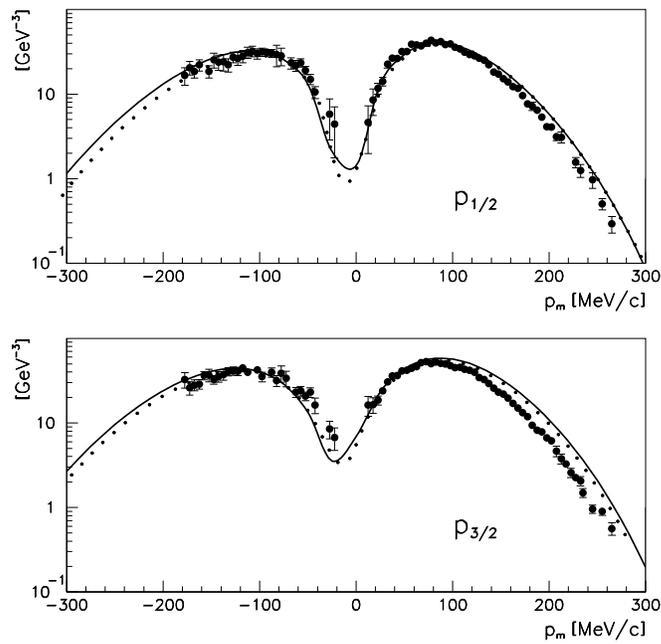}
\vspace{2mm}\caption{Reduced cross sections of  the
$^{16}$O($e,e'p$) reaction as a function of the missing momentum 
$p_{\mathrm m}$ for the transitions to the $1/2^{-}$ ground state and to the 
first $3/2^{-}$ excited state of $^{15}$N in the same kinematics as in 
Fig.~1. The solid lines give the RDWIA result [36] the dotted lines the
nonrelativistic result. The optical potential is from ref. [39] and the 
bound state wave functions from ref. [40] (from ref. [36]).
\label{fig:fig3}
}
\end{center}
\end{figure}

An example is shown in Fig.~\ref{fig:fig3}, for the $^{16}$O($e,e'p$) reaction 
in comparison with the NIKHEF data \cite{Leuschner}. Only small differences are 
found between the two calculations in this kinematics. The reduction 
(spectroscopic) factor applied to the calculated reduced cross sections in 
order to reproduce  the size of the experimental results is 0.7 for RDWIA and 
0.65 for DWIA, for both the transitions, which confirms that somewhat higher 
spectroscopic factors are obtained in RDWIA. 

The systematic investigation carried out in ref. \cite{Meucci} indicates that 
relativistic effects increase with the energy and in particular with the energy 
of the outgoing proton. The DWIA
approach can be used with enough confidence at the energies around 100 MeV of
previous ($e,e'p$) experiments, and, with some caution, up to about 200 MeV. 
This confirms the validity of the analyses carried out with DWEEPY at lower 
energies. A fully relativistic calculation is anyhow convenient at 200 MeV 
and necessary above 300~MeV. Thus, RDWIA must be used in comparison with the 
recent data from JLab \cite{Gao}, at $Q^2$ = 0.8 (GeV/$c$)$^2$ and $T'_1 = 433$ 
MeV. Here the RDWIA model gives an excellent description of data keeping the 
same spectroscopic factor (0.7) extracted in the comparison with the NIKHEF 
data of Fig.~\ref{fig:fig1} \cite{Meucci}.

\subsection{Conclusions}

The results of the study of $NN$ correlations in the ($e,e'p$) reaction on 
complex nuclei can be summarized as follows. The s.f. account for the depletion 
of the quasi-hole states produced by $NN$ correlations. The depletion found in 
the DWIA analyses is $\sim 30-40$\%. The s.f. are usually extracted from the 
comparison between data and DWIA results and can be affected by all the 
uncertainties of the model. Theoretical investigations within different 
correlation methods indicate that only a few percent of the depletion is due to 
SRC (see also \cite{FC,MACL}). When TC are added to SRC the depletion amounts to 
$\sim 10$\%, at most $\sim 15$\% in heavy nuclei. Further depletion is given by 
LRC. A full and consistent calculation of the OF and of the s.f. including SRC, 
TC and LRC is still unavailable. 

If we want to study more specifically SRC, we have seen that they account for
only a small part of the depletion of the quasi-hole states. This depletion
is compensated by the admixture of high-momentum components in the nuclear wave
function. Thus, one might think to investigate SRC studying the high-momentum
components of the s.p. wave functions in exclusive one nucleon knockout
experiments. Indeed we have found large differences for the cross sections
calculated with the different OF at high values of the missing momentum. It is 
not clear, however, if these differences are due to correlations or to the 
different methods used in the calculations of the OBDM. Microscopic calculations 
of the momentum distribution \cite{MP} give indeed a strong enhancement of the 
high-momentum components due to SRC, but this enhancement shows up at large 
values of $E_{\mathrm m}$. In exclusive ($e,e'p$) experiments one does not 
measure the whole momentum distribution, but the spectral function at the 
energy corresponding to the specific final state that is considered. In general 
low-lying discrete states of the residual nucleus are considered, corresponding 
to low values of the energy, while the missing strength due to SRC is found at 
high values of the momentum but also at large values of the excitation energy, 
well above the continuum threshold, where other competing processes are present. 
This makes a clear-cut identification of SRC in ($e,e'p$) very difficult.

This identification appears possible in two-nucleon knockout reactions. Here 
particular situations can be envisaged where the knockout of the two nucleons is
entirely due to correlations. These situations appear very well suited to study 
SRC. 
 
\section{Two-Nucleon Knockout}

Since a long time electromagnetically induced two-nucleon knockout reactions 
have been devised as the preferential tool for investigating SRC. In fact, 
direct insight into SRC can be obtained from the situation where the 
electromagnetic probe hits, through a one-body current, either nucleon of a 
correlated pair and both nucleons are then ejected from the nucleus. This 
process is entirely due to correlations. But two nucleons can also and 
naturally be ejected by two-body currents due to meson exchanges and $\Delta$ 
isobar excitations.  These two competing processes, both produced by the 
exchange of mesons between nucleons, require a careful and consistent treatment. 
Their role and relevance, however, is different in different reactions and 
kinematics. It is thus possible, with the help of theoretical predictions, to 
envisage appropriate situations where various specific effects can be 
disentangled and separately investigated.   

Interesting and complementary information is available from electron and
photon-induced reactions, but the electron probe is preferable to study SRC. In 
fact, two-body currents predominantly contribute to the transverse components of 
the nuclear response. Only these components are present in photon-induced 
reactions that appear thus generally dominated by two-body currents. Also the 
longitudinal component, dominated by correlations, is present in 
electron-induced reactions. The possibility of independently varying the energy 
and momentum transfer of the exchanged virtual photon allows one to select 
kinematics where the longitudinal response and thus SRC are dominant. 

A combined study of $pp$ and $np$ knockout is needed for a complete information. 
Correlations are different in $pp$ and $np$ pairs. They are stronger in $np$ 
pairs and thus in $np$ knockout due to the tensor force, that is predominantly 
present in the wave function of a $np$ pair. But also two-body currents are much 
more important in $np$ knockout, while they are strongly suppressed in $pp$ 
knockout, where the charge-exchange terms of the two-body current do not 
contribute. Therefore, the ($e,e'pp$) reaction was devised as the preferential 
process for studying SRC in nuclei. It is however clear that, since different
effects can be emphasized in suitable conditions for different reactions, 
a combined study of $pp$ and $np$ knockout induced by real and virtual photons 
is needed to unravel the different contributions and obtain clear and complete 
information on $NN$ correlations. 

Exclusive reactions, for transitions to specific discrete eigenstates of the
residual nucleus, are of particular interest for this study. One of the main
results of the theoretical investigation is the selectivity of exclusive 
reactions involving different final states that can be differently affected by 
one-body and two-body currents \cite{giu98,pn}. Thus, the experimental 
resolution of specific final states may act as a filter to disentangle the two 
reaction processes. $^{16}$O is a suitable target for this study, due to the 
presence of discrete low-lying states in the experimental spectrum of $^{14}$C 
and $^{14}$N well separated in energy. From this point of  view, $^{16}$O is 
better than a light nucleus, which lacks specific final states. 

\subsection{The Theoretical Framework}

The theoretical framework for two-nucleon knockout is formally similar to the 
model for one-nucleon knockout outilned in Sec. II. The transition matrix 
elements in Eq.~(\ref{eq:jm}), whose bilinear combinations give the components
of the hadron tensor $W^{\mu\nu}$, represent the basic ingredients of the
calculation. The model \cite{Oxford,GP91} is still based on the two assumptions 
of an exclusive process, for the transiton to a discrete eigenstate of the 
residual nucleus, and of the DKO mechanism, but a different final nuclear state 
must be considered, with two outgoing nucleons and the residual nucleus, and 
the nuclear-current operator is the sum of a one-body and a two-body part,
corresponding to the two competing reaction processes already mentioned. The 
two-body current includes terms due to the lowest order diagrams with 
one-pion exchange, namely seagull, pion-in-flight and diagrams with 
intermediate $\Delta$ isobar configurations \cite{GP98}. All these terms 
contribute to $pn$ knockout while only the non charge-exchange terms in the 
$\Delta$ current operator contribute to $pp$ knockout.  
  
The matrix elements of Eq.~(\ref{eq:jm}) can be written as
\begin{equation}
J^{\mu}(\q) = \int \psi_{\mathrm{f}}^{*}(\r_1,\r_2) J^{\mu}(\r,\r_1,\r_2)
\psi_{\mathrm{i}}(\r_1,\r_2){\mathrm{e}}^{\,{\mathrm{i}} {\footnotesize {\q}} 
\cdot {\footnotesize {\r}}} {\mathrm d}\r {\mathrm d}\r_1 {\mathrm d}\r_2 ,
\label{eq:j2}
\end{equation} 
where the two-nucleon overlap integral $\psi_{\mathrm{i}}$ and the two-nucleon
scattering state $\psi_{\mathrm{f}}$ are consistently derived from an 
energy-dependent non-Hermitean Feshbach-type Hamiltonian for the considered 
final state of the residual nucleus. In practice, since it would be extremely
difficult to achieve this consistency, the treatment of initial and final 
states proceeds separately with different approximations.

In the scattering state the mutual interaction between the two outgoing nucleons 
is neglected and only the interaction of each of the outgoing nucleons with the 
residual nucleus is considered by means of a phenomenological optical potential.

For the $^{16}$O($e,e'pp$)$^{14}$C reaction the two-nucleon overlap functions 
are taken from the calculation of the spectral function \cite{giu98,SF},
where both LRC and SRC are included. A two-step procedure has been applied in
this calculation where LRC and SRC are treated in a separate but consistent way.
The calculation of LRC is performed in a SM space large enough to incorporate 
the corresponding collective features which influence the pair removal 
amplitudes. The s.p. propagators used for this dressed Random Phase 
Approximation (RPA) description of the two-particle propagator also include the 
effect of both LRC and SRC. This yields s.f. for low-lying states of $^{15}$N
which represent the closest agreement with ($e,e'p$)  data to date 
\cite{Geurts}. In the second step that part of the pair removal 
amplitudes which describes the relative motion of the pair is supplemented by 
defect functions obtained from the same G-matrix which is also used as the 
effective interaction in the RPA calculation.

The two-nucleon OF for a discrete final state of $^{14}$C, with angular 
momentum quantum numbers $JM$, is expressed in terms of a combination of 
relative and c.m. wave functions \cite{giu98}. The combination coefficients 
contain contributions from a SM space which includes the $0s$ up to the $1p0f$ 
shells. The c.m. radial wave function is that of a harmonic oscillator
(h.o.). SRC are included in the radial wave function $\phi$ of relative motion 
through a defect function defined by the difference between $\phi$ and the 
uncorrelated relative h.o. wave function.  
These defect wave functions depend on the quantum numbers of the 
relative motion. Thus, SRC depend  on the relative state and, since different
components of relative and c.m. motion contribute to each transition, the
role of SRC can be different for different final states. 

\subsection{Evidence for SRC in the $^{16}$O($e,e'pp$)$^{14}$C Reaction}

\begin{figure}[ht]
\begin{center}
\includegraphics[height=100mm,width=120mm]{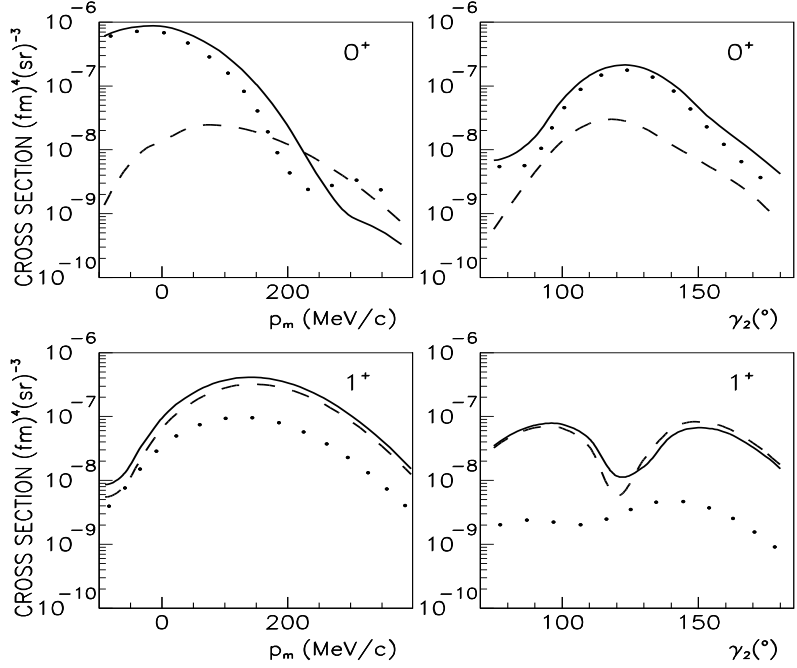}
\vspace{2mm}
\caption{The differential cross section of the reaction 
$^{16}$O($e,e'pp$)$^{14}$C for the transitions to the $0^+$ ground state and to 
the $1^+$ state at 11.31 MeV. In the left panels a super-parallel kinematics is 
considered with $E_0=855$ MeV, $\omega=215$ MeV and $q=316$ MeV/$c$. Positive 
(negative) values of $p_{\mathrm{m}}$ refer to situations where 
$\p_{\mathrm{m}}$ is parallel (anti-parallel) to $\q$. In the right panels 
$E_0=584$ MeV, $\omega=212$ MeV, $q=300$ MeV/$c$, $T'_1=137$ MeV and 
$\gamma_1= -30^{\rm{o}}$, on the opposite side of the outgoing electron 
with respect to the momentum transfer. The defect functions for the 
Bonn-A $NN$ potential and the optical potential of ref. [24] are 
used. Separate contributions of the one-body and the two-body $\Delta$ current 
are shown by the dotted and dashed lines, respectively. The solid curves give 
the final result.
\label{fig:fig4}
}
\end{center}
\end{figure}

A numerical example is shown in Fig.~\ref{fig:fig4}, where the cross sections of
the $^{16}$O($e,e'pp$)$^{14}$C reaction are displayed for the transitions to 
the $0^+$ ground state and to the $1^+$ state at 11.31 MeV. Results are shown 
for two kinematical settings considered in the experiments performed at 
NIKHEF \cite{Ronald,Gerco} and MAMI \cite{Rosner}. 

Different components of relative and c.m. motion contribute to the two final 
states \cite{giu98}:  $^1S_0$ and $^3P_1$ relative waves (the notation 
$^{2S+1}l_j$, for $l = S,P,D,$  is used here for the relative states), which 
are combined with a c.m. orbital angular momentum $L=0$ 
and $1$, respectively, for the $0^+$ state, and $^3P_0$, $^3P_1$, $^3P_2$, all 
combined with $L=1$, for the $1^+$ state. The value of $L$ determines the shape 
of the recoil-momentum distribution. Indeed in Fig.~4 for the  $1^+$ state, 
where only components with $L=1$ are present, the momentum distributions have a 
typical $p$-wave shape, while the $s$-wave shape obtained for the $0^+$ state 
indicates that in the two considered kinematics the cross section is dominated 
by the component with $L=0$ and thus by $^1S_0$ $pp$ knockout. The component 
with $L=1$, due to $^3P_1$, becomes meaningful only at large values of 
$p_{\mathrm{m}}$, where the contribution of the $s$ wave gets lower.

The comparison between correlated and uncorrelated relative wave 
functions \cite{giu98,SF} indicates that SRC play a different role in 
different relative states: they are quite strong for the $^1S_0$ state and much 
weaker for $^3P$ states. Moreover, also the role of the isobar current is 
strongly reduced for $^1S_0$ $pp$ knockout, since there the generally dominant 
contribution of that current, due to the magnetic dipole  
$NN \leftrightarrow N\Delta$ transition, is suppressed \cite{GP98,delta}. Thus,
the role of SRC is emphasized in $^1S_0$ knockout, while the role of the 
$\Delta$ current is emphasized in $^3P$ knockout. This explains the different
role of the two reaction processes for the two final states in 
Fig.~\ref{fig:fig4}: the transition to the $1^+$ state is dominated by the 
two-body current, while for the $0^+$ state, where $^1S_0$ $pp$ knockout plays 
the main role, the cross section is dominated by the one-body current and thus 
my SRC. 

The final result is determined by all the ingredients of the model, 
but it is clear that the two reaction processes play a different role for the 
two final states. Thus, the experimental resolution of different states may act 
as a filter to disentangle and separately investigate the contributions due to 
SRC and two-body currents.
 
Data have confirmed the predictions of this model. A reasonable and in some 
cases an excellent agreement with the available data \cite{Ronald,Gerco,Rosner} 
has been obtained. The comparison has clearly shown the validity of the DKO 
mechanism for transitions leading to the lowest-lying states of $^{14}$C and 
has confirmed the predicted selectivity of the exclusive reaction involving 
discrete final states, which are differently affected by SRC and two-body 
currents. In particular, clear evidence for SRC has been obtained for the 
transition to the ground state \cite{Ronald}. 

This important result means that two-nucleon knockout reactions can be used to
study and hopefully determine SRC. More theoretical and experimental work is
however needed for this study. More data are expected from MAMI for all the 
exclusive $^{16}$O($e,e'pp$)$^{14}$C, $^{16}$O($e,e'np$)$^{14}$N, 
$^{16}$O($\gamma,pp$)$^{14}$C, and $^{16}$O($\gamma,pn$)$^{14}$N knockout 
reactions \cite{MAMI}. A combined study of different reactions is needed for a 
complete information on $NN$ correlations.

Good opportunities to increase the richness of information available from
two-nucleon knockout reactions  are also offered by polarization 
measurements \cite{pol}. Reactions with polarized particles give access to a 
larger number of observables, hidden in the unpolarized case, whose 
determination can impose more severe constraints on theoretical models. 
Thus, a combined analysis of cross sections and polarization observables would 
make possible it to disentangle the different contributions and shed light on 
the genuine nature of correlations in nuclei.

\section*{}

I want to thank all the colleagues who contributed to this work, in 
particular A.N. Antonov, M.K. Gaidarov, A. Meucci and F.D. Pacati.

\begin{table}
\bigskip
\caption[Table I]{Spectroscopic factors for the $^{16}$O($e,e'p$)
knockout reaction leading to the $1/2^{-}$ ground state and to the $3/2^{-}$
excited state of $^{15}$N. Column I gives the s.f. deduced
from the calculations with different OBDM of $^{16}$O; II gives the additional
reduction factors determined through a comparison between the ($e,e'p$) data of
ref. \cite{Leuschner} and the reduced cross sections calculated in DWIA with
the different overlap functions; III gives the total s.f.
obtained from the product of the factors in columns I and II.}
\bigskip
\begin{center}
\begin{tabular}{cccccccc}
\hline\hline
&  \multicolumn{3}{c}{$1p_{1/2}$} & &  \multicolumn{3}{c}{$1p_{3/2}$} \\
\cline{2-4} \cline{6-8}
OBDM & I   & II  & III & & I   & II  & III \\
\hline
HF               & 1.000 & 0.750 & 0.750 & &  1.000 & 0.550 & 0.550\\
JCM \cite{Sto96} & 0.953 & 0.825 & 0.786 & &  0.953 & 0.600 & 0.572\\
CBF \cite{Vn97}  & 0.912 & 0.850 & 0.775 & &  0.909 & 0.780 & 0.709\\
CBF \cite{Sa96}  & 0.981 & 0.900 & 0.883 & &  0.981 & 0.600 & 0.589\\
GFM \cite{Po96}  & 0.905 & 0.800 & 0.724 & &  0.915 & 0.625 & 0.572\\
GCM \cite{Chr86} & 0.988 & 0.700 & 0.692 & &  0.988 & 0.500 & 0.494\\
\hline\hline
\end{tabular}
\end{center}
\end{table}

\end{document}